\documentclass{vospwks2007}

\title{PopStar: A new grid of Evolutionary Synthesis Models in the Virtual Observatory}
\author{Marisa Garc\'{\i}a-Vargas}
\affil{FRACTAL SLNE. Castillo de Belmonte 1,  b5, bajo A. E-28232 Las Rozas de Madrid (Spain)}
\author{Mercedes Moll\'{a}}
\affil{CIEMAT. Avda. Complutense-22. E-28040 Madrid (Spain)}
\author{Alessandro Bressan}
\affil{INAF. Osservatorio Astronomico di Padova. Viccolo dell' Osservatorio-5. 35122 Padova (Italy)}
\author{Pedro G\'{o}mez-Alvarez}
\affil{ESAC. Villafranca del Castillo, P.O. Box 50727. E-28080 Madrid (Spain)}

\usepackage{graphicx}
\begin{document}


\maketitle

\begin{abstract}
We present a new set of theoretical evolutionary synthesis models, PopStar. This grid of
Single Stellar Populations covers a wide range in both, age and metallicity. The models use
the most recent evolutionary tracks together with the use of new NLTE atmosphere models for the hot
stars (O, B, WR, post-AGB stars, planetary nebulae) that dominate the stellar cluster's ionizing spectra.
The results of the models in VO format can be used through VOSpec.
\end{abstract}

\section{Models Description}

We have used the synthesis code by \citet{gmb:1998}, updated by
\citet{mgv:2000} and newly revised now. The basic grid is composed by
Single Stellar Populations (SSP) for five different IMFs: two based on Salpeter power law \citep{sal:1955}, with
mass range 0.85 - 120 M$_{\odot}$ and 0.15 - 100 M$_{\odot}$ respectively, and those of \citet{fpp:1990},
\citet{kro:2002} and \citet{cha:2003} functions, with masses between 0.15 and 100 M$_{\odot}$.
These models do not include binaries either mass segregation. The isochrones are those from \citet{bgs:1998} for 6 different metallicities: Z $=$ 0.0004, 0.001, 0.004, 0.008, 0.02 and 0.05.
The age coverage is from $log{t}=$ 5.00 to 10.30, with a variable time resolution which is
$\Delta(log{t})=0.01$ in the youngest stellar ages. The WC and WN stars are identified in the isochrones according
to their surface abundances.

\begin{figure*}
\centering
\includegraphics[angle=90,scale=0.50]{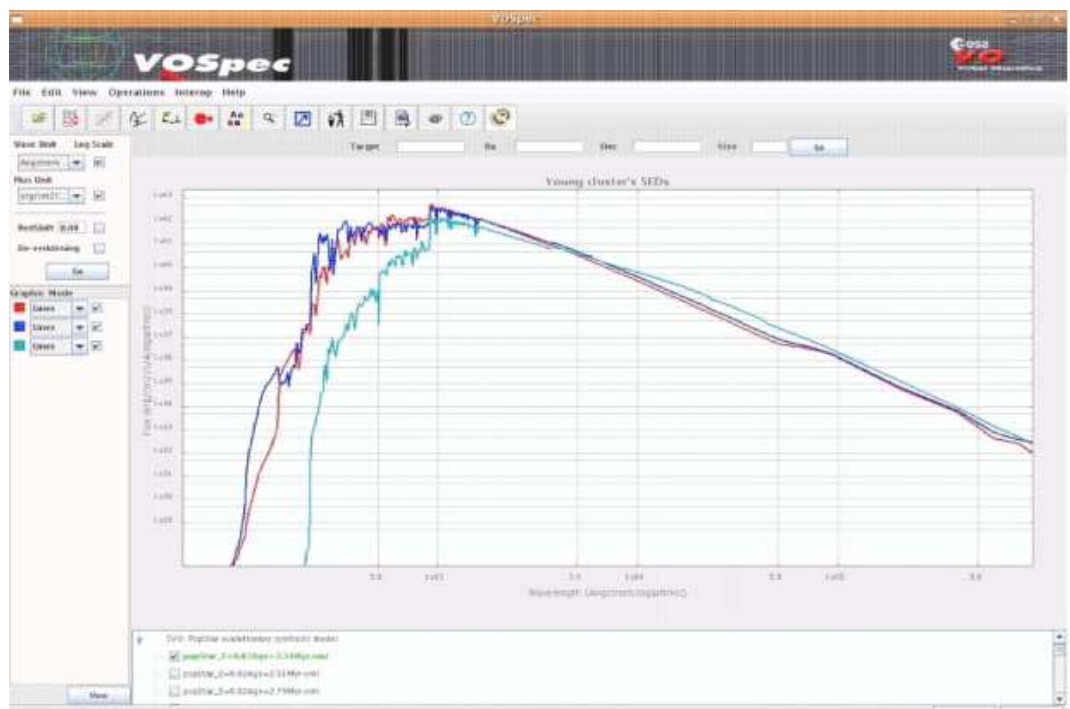}
\caption{PopStar SEDs for young clusters displayed through the VOSpec interface \label{fig:single}}
\end{figure*}

The atmosphere models are from \citet{lcb:1997} with an excellent coverage
in effective temperature, gravity and metallicities, for stars with T$_{eff}$ $ \leq 25 000$ K.
For O, B and WR we have taken the NLTE blanketted models by \citet{snc:2002} at 5 different metallicities.
There are 110 models for O-B stars, calculated by \citet{phl:2001}, with $25 000$ K $ <$
T$_{eff}$ $\leq 51 500$ K and $2.95 \leq \log{g} \leq 4.00$, and 120 for WR stars (60 WN and 60 WC), from \citep{hm:1998},
with 30 000 K $\leq T^{*}\leq 120 000$ K and $1.3R_{\odot}\leq R^{*}\leq 20.3 R_{\odot}$ for WN,
and with $ 40 000 K \leq\ T^{*} \leq 140 000$ K and $ 0.8R_{\odot}\leq\ R^{*}\leq 9.3 R_{\odot}$ for WC. T$^{*}$
and R$^{*}$ are the temperature and the radius at a Roseland optical depth of 10. To assign an appropriate model
to each WR star, we use the relationships among opacity, mass loss and velocity wind:
$d\tau=-\kappa(r)\rho(r)dr$, where $\kappa(r)=-0.2(1+X_{\rm S})$ and
$X_{\rm S}$, the H surface abundance, is taken as 0.2 for WN and 0 for WC. The mass loss is: $dM/dt =4\pi
r^{2}\rho(r)v(r)dr$ with $v(r)=v_{\infty}(1-R_{S}/r)^{\beta}$ taking $\beta=1$.  Integrating these equations
we find R$^{*}$ from the $R_{S}$ of isochrones selecting the closer atmosphere model to this R$^{*}$.
Above 54ev emergent flux is determined by the wind density as a function of metallicity and in this set of
models the hardness of ionizing radiation decreases, producing a lower flux of HeI continuum,
for $Z > 0.4Z_{\odot}$ and t $< 7$ Myr. HeII 4686 is very low and HeI 5786 / H$_\beta$ decreases when metallicity increases due to line blanketing.

For post-AGB and Planetary Nebulae with T$_{eff}$ ranging from 50 000 K to 220 000 K we take the NLTE models by \citet{rau:2003} with values of T$_{eff}$  between 50 000 K and 190 000 K and logg between 5.00 and 8.00. These models include all elements from H to Ni. For higher temperatures we use black bodies.

\section{VO Products}

We provide the SED in the VO standard: $\lambda$ (in \AA) and L$_{\lambda}$ (in ergs$^{-1}$\AA$^{-1}$)
Spectra can be selected by IMF type, age and Z and can be managed with the VOSpec tool.
We have produced the HR diagram files covering the whole grid in age and metallicity. At the moment, there is not a specific VO access protocol for such files. However, we have produced VO tables that can be used with some VO tools like Topcat. A detailed description of the models and the products is available at the webpage http://www.fractal-es.com/SEDmod.htm/, where the files can be downloaded.

\section{Conclusions}

We have computed a new grid of SEDs with updated isochrones and atmosphere models in a wide range of
age and metallicity. The use of NLTE blanketed models produce SEDs with less hard ionizing photons, able to explain
the observed emission line spectrum in low excitation high metallicity H{\sc ii} regions without changing the IMF
parameters. Previous models \citep{gbd:1995} could not explain these observations without eliminating the
more massive stars artificially (with a flatter IMF, a lower value of the upper mass limit or using the standard IMF
but considering the large clusters divided in small less massive sub-cluster with a low probability of forming very
massive stars). H-R Diagrams are available as VO tables. The resulting SEDs are in VO format accessible through SSAP protocol and can be managed with VOSpec.

\section*{Acknowledgments}

M. Moll\'{a} and M. Garc\'{\i}a-Vargas are researchers of the project Estallidos (AyA2004-08260-C03).
The authors thank Pedro Osuna, Jes\'{u}s Salgado, Isa Barbarisi (ESA-VO), Enrique Solano and Miguel Cervi\~{n}o
(SVO) their support in VO issues. The Spanish Virtual Observatory is supported from the Spanish MCyT through
grants AyA2005-04286, AyA2005-24102-E.


\begin{thebibliography}{}

\bibitem[Bressan et~al.(1998)]{bgs:1998} Bressan A., Granato G.L., \& Silva L. 1998, A\&A, 232, 135

\bibitem[Chabrier(2003)]{cha:2003} Chabrier, G.\ 2003, ApJL, 586, L133

\bibitem[Ferrini et~al.(1990)]{fpp:1990} Ferrini F., Penco U., and Palla F., A\&A, 231, 391

\bibitem[Garc\'{\i}a-Vargas et~al.(1995)]{gbd:1995}
Garc\'{\i}a-Vargas M.L., Bressan A., \& D\'{\i}az A.I. 1995, A\&AS,112, 13

\bibitem[Garc\'{\i}a-Vargas et~al.(1998)]{gmb:1998}
Garc\'{\i}a-Vargas M.L., Moll\'{a} M., \& Bressan A. 1998, A\&AS,130, 513

\bibitem[Hillier \& Miller(1998)]{hm:1998}
Hillier, D.J., \& Miller, D.L. 1998, ApJ,, 496, 407

\bibitem[Kroupa(2002)]{kro:2002}
Kroupa, P.\ 2002, Science, 295, 82

\bibitem[Lejeune et~al.(1997)]{lcb:1997}
Lejeune Th., Cuisinier F.,\& Buser R. 1997, A\&AS, 125, 229

\bibitem[Moll\'{a}, \& Garc\'{\i}a-Vargas(2000)]{mgv:2000}
Moll\'{a}, M., \& Garc\'{\i}a-Vargas, M.L. 2000, A\&A, 359,18

\bibitem[Pauldrach et~al.(2001)]{phl:2001}
Pauldrach A., Hoffmann T.L., \& Lennon M. 2001, A\&A. 375, 161

\bibitem[Rauch(2003)]{rau:2003}
Rauch, T. 2003, A\&A 403, 709

\bibitem[Salpeter(1955)]{sal:1955}
Salpeter, E.E. 1955, ApJ, 121, 161

\bibitem[Smith et~al.(2002)]{snc:2002}
Smith L., Norris R., \& Crowther P. 2002, MNRAS, 337, 1309

\end{thebibliography}
\end{document}